\documentclass[]{elsarticle} 
\usepackage{hyperref} 

\usepackage{amssymb}
\usepackage{amsmath}
\usepackage{epsfig}

\newcommand{\be}{\begin{equation}}
\newcommand{\ee}{\end{equation}}
\newcommand{\benn}{\begin{displaymath}}
\newcommand{\eenn}{\end{displaymath}}
\newcommand{\ba}{\begin{eqnarray}}
\newcommand{\ea}{\end{eqnarray}}
\newcommand{\nn}{\nonumber}

\newcommand{\wt}{\widetilde}
\renewcommand{\vec}[1]{\mbox{\boldmath $#1$}}

\renewcommand{\d}[1]{{\rm d}#1}
\newcommand{\half}{\mbox{$\frac{1}{2}$}}

\newcommand{\reyn}{{\rm Re}}

\journal{Journal of Biomechanics}

\begin{document} 
\begin{frontmatter}

\title{The effect of finger spreading on drag of the hand in human swimming}

\cortext[whomtowrite]{Corresponding authors}
\author[tue]{Josje van Houwelingen\corref{whomtowrite}}
\ead{j.v.houwelingen@tue.nl}
\author[tue]{Dennis H.J. Willemsen}
\author[tue]{Rudie P.J. Kunnen}
\author[tue]{GertJan F. van Heijst}
\author[tud]{Ernst Jan Grift}
\author[tud]{Wim Paul Breugem}
\author[tud]{Rene Delfos}
\author[tud]{Jerry Westerweel}
\author[tue]{Herman J.H. Clercx}
\author[tue,tud]{Willem van de Water\corref{whomtowrite}}
\ead{w.v.d.water@tue.nl}
  
\address[tue]{Department of Applied Physics, 
  Eindhoven University of Technology 
  and J.M. Burgers Centre for Fluid Dynamics,
  Postbus 513, 5600 MB Eindhoven,
  The Netherlands. 
}
\address[tud]{
  Laboratory for Aero and Hydrodynamics, 
  Delft University of Technology
  and J.M. Burgers Centre for Fluid Dynamics,
  2628 CD Delft,
  The Netherlands
}

\begin{abstract}
The effect of finger spreading on hydrodynamic drag in swimming is
studied both with a numerical simulation and with laboratory
experiments. Both approaches are based on the exact same 3D model of
the hand with attached forearm. The virtual version of the hand with
forearm was implemented in a numerical code by means of an immersed
boundary method and the physical version was studied in a wind tunnel
experiment. An enhancement of the drag coefficient of 2 and 5\%
compared to the case with closed fingers was found for the numerical
simulation and experiment, respectively. A 5 and 8\% favourable
effect on the (dimensionless) force moment at an optimal finger
spreading of $10^\circ$ was found, which indicates that the
difference is more outspoken in the force moment. Also an analytical
model is proposed, using scaling arguments similar to the Betz
actuator disk model, to explain the drag coefficient as a function of
finger spacing.  
\end{abstract}


\end{frontmatter}


\section{Introduction} 
%

%
A tantalizing question in front-crawl swimming is whether the stroke
efficiency during the propulsive phase depends on spreading the fingers.
Several studies exist that suggest a small increase of the drag of
the hand in that case.  The idea is that, although the projected area
of the hand remains the same, a slight opening of the fingers still
provides enough obstruction to the flow, which must be forced between
the fingers.  Such an increase of the drag increases thrust, or gives
the same thrust at lower hand velocities, both leading to enhanced
swimming efficiency. For top swimmers, this may provide a competitive
edge.  

\noindent Since the effect is so small, its measurement or computation using
numerical simulation of the flow is a challenge.  Several studies
exist that involve either experiments or numerical simulations, which
makes a comparison between experimental and numerical results
difficult.
A good comparison requires a well-defined model for the hand (and
forearm), combined experimental measurements and numerical
computations, and clear definitions of hydrodynamic quantities to be
used. For example, to quantify the drag $F_D$, the quantity of
interest is the drag coefficient, $C_D = F_D / \half \rho A U^2$,
where $A$ is the area of the hand projected perpendicular onto the direction of hand
motion, $\rho$ the density of the fluid and $U$ the incoming flow
velocity.  

\noindent Despite differences in used models and parameter settings it is
worthwhile to provide a brief overview of the main conclusions in the
literature so far. \citet{Schleihauf1979} was the first to study the
influence of the hand orientation and finger spacing on the
hydrodynamics of swimming by conducting water channel experiments.
Although marginal differences, no advantage in drag and lift
coefficients for a 1.27~cm or 0.64~cm 
finger spread was found compared to closed fingers. 
\citet{Sidelnik2006} performed an experiment based on an unsteady
approach. A robotic arm model with closed fingers and a $10^{\circ}$
finger spread, mimicking sculling motion, was towed through a water
tank. Higher propulsive forces were obtained with spread fingers.  
In a numerical study, \citet{Minetti2009} reported a $C_D$ value of
0.52 and found a 8.8\% drag increase compared to closed fingers for a
$13^\circ$ finger spacing. The size of the wake and strength of the
vortices seem to determine the results \cite{Minetti2009}. The widest
wake was found with optimal finger spreading. With closed fingers,
larger vortices were present in the wake, while at optimal spacing
the creation of such vortices was constrained by the presence of jets
between the fingers. 
Also \citet{Marinho2010} computed the $C_D$ for three finger
spreadings and found $C_D = 1.1$ with a 5\% increase of the drag
coefficient for a small finger spreading compared to the case of
closed fingers. They proposed that a barrier of turbulent flow is
formed between the fingers. Note that the values of $C_D$ by
\citet{Minetti2009} are quite different, possibly because of 
different sized hand and arm models (section forearm included) 
resulting in a different surface $A$. Other small discrepancies can 
be assigned to shape differences in hand and arm models (thumb 
position, definition finger spacing).
%
\citet{Lorente2012} explained the effect of finger spreading by
representing the hand as an array of four cylinders. They argue that
optimal spreading is the consequence of overlapping (laminar)
boundary layers of individual fingers. However, the Reynolds number
of their simulations is small ($\reyn = 100$), where the effect of
optimal spreading (a 28\% increase) appears unrealistically large.
Finally, in their steady state simulations for several orientations
and hand models, \citet{Bilinauskaite2013} did not find an increase
of the drag force or drag coefficient when using slightly opened
fingers. However, they conclude that the maximal values of local
pressure for spread fingers indicate a higher pressure force and thus
a higher drag force \cite{Minetti2009,Marinho2010}. 
Besides finger spread, a variety of fluid dynamic aspects of swimming
has recently been reviewed by \citet{Wei2014} and \citet{Takagi2016} while \citet{Houwelingen2016} focused on the hydrodynamics of the swimmer's hands. 

\noindent The character of the flow is quantified by the Reynolds number,
$\reyn = U W / \nu$, where $W$ is the width of the hand palm or
finger, and $\nu$ the kinematic viscosity. Taking the relative
velocity of the hand with respect to the water $U \sim 1$~m/s and the
width of a finger $\sim$ 0.01~m, gives $\reyn \approx 10^4$. For a
hand palm with width $\sim$ 0.1~m, $\reyn$ is an order of magnitude
larger. 
For example, the thickness of (laminar) boundary layers, denoted by
$\delta$, around an object of size $W$ is approximately $\delta
\propto 5 \reyn^{-1/2} W$ \cite{Prandtl1956}. With $W = 0.01$~m, the
typical boundary layer thickness around the finger is $\delta \approx
5\times 10^{-4}$~m. If the boundary layer is indeed of importance in
describing this problem, a numerical simulation will be a challenge
as a prohibitive amount of grid cells has to be used to completely
resolve the flow, also in the boundary layers.

\noindent In this study we follow the unique approach where exactly the same
hands were used both in numerical simulations and wind tunnel
experiments, at the same value of the Reynolds number. The
simulations are performed with an immersed boundary method, which is
in favour for complex shaped (and moving) bodies. 
A similar computational technique has been used in a study on the
underwater kick and different arm pull styles \cite{Loebbecke2009,Loebbecke2012}, but also in the analysis of flying and swimming in
nature \cite{Mittal2006}, showing the tremendous potential of this
approach.  Five different hands were used with five different angles
$\theta$ between the fingers: the closed hand with $\theta = 0$ and
increasingly spread fingers with $\theta = 5, 10, 15$ and 20 degrees
(see Fig.\ \ref{fig.setup} (a)), respectively. Apart from mean flow
quantities, such as drag of the entire arm and the drag moment around
the elbow joint, we have also studied the fluctuations in the forces.
Finally, by analysing a Betz-type actuator model of the hand with
variable finger spreading, for the first time a simple moment-balance
explanation of the observed drag optimum is provided.


\section{Virtual and real hands}
\begin{figure*}[t]
\centerline{\epsfxsize = 120 mm \epsffile{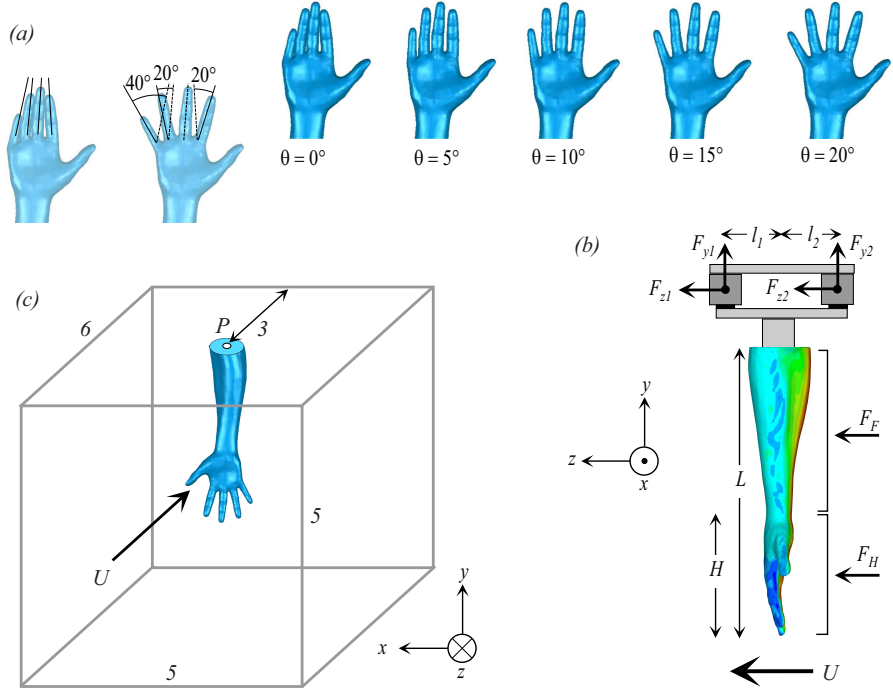}}
\caption{
  Panel (a) shows the five hands used.  The finger spreads are
  0, 5, 10, 15 and $20^\circ$. The angle of each finger is defined as
  the angle relative to the $0^\circ$ position. For example, for the
  $20^\circ$ model the index finger and ring finger are positioned
  $20^\circ$ outward compared to their position in the $0^\circ$
  model, whereas the little finger is bent $40^\circ$ outward. These
  angles roughly correspond to a spacing between the finger tip side
  edges of 0, 7.5, 15, 20 and $25-30\:{\rm mm}$, respectively.  
  Panel (b) is a schematic view of the experimental setup in the wind
  tunnel. Two force sensors are used in tandem configuration, which
  gives access to forces $F_{x,y,z}$ and force moment $M_x$.  The
  total drag force is $F_D = F_{z_1} + F_{z_2}$, the drag moment
  is $M_x = l_2 F_{y_2} - l_1 F_{y_1}$.  The forces on
  the forearm and hand are $F_F$ and $F_H$, respectively. 
  Panel (c) shows the computational domain.  Force moments are
  computed with respect to the point $P$.} 
\label{fig.setup}
\end{figure*}

The hands were made using the public domain software {\em Make Human}
\cite{Makehuman2016}, which contains a virtual 3D model of a human,
including (natural) joints. The program allowed to select the hand
and arm and spread the fingers by rotating them around the joints, as
is illustrated in Fig. \ \ref{fig.setup}(a). In combination with 3D
meshing tools 
the hand was turned into an STL format which was used in the
simulations and 3D printed for the wind tunnel. In order to study the
effect of the finger spreading alone, the thumb had a fixed position.
Changing the thumb abduction would have changed the projected area of
the hand and would have influenced the drag and lift characteristics
of the hand \cite{Schleihauf1979,Bilinauskaite2013,Takagi2001,Marinho2009}.

\noindent The hand and forearm were digitized with $7.5\times 10^4$ triangular
surface elements. The virtual arms, with a hand palm width, length and frontal surface of 0.59, 3.1 and 1.57, in dimensionless simulations units respectively,
were used in the numerical simulations. For the wind tunnel
tests the same files were used to print physical arms with 3D
printing technology, with a hand palm width, length and frontal
surface of 0.096~m, 0.507~m and 0.042~m$^2$, respectively.

\noindent The numerical simulations were performed using the immersed boundary
method (IBM) \cite{Iaccarino2003}. In this IBM, the Navier-Stokes equations (NS) for an
incompressible flow are solved by an integration scheme based on the
fractional step method \cite{Kim1985}. A finite difference scheme is
used for discretization of the derivatives \cite{Kim1985,Verzicco1996}. IBM has the great advantage that complex boundaries
can be adapted readily, without the burden of body-adapted grid
generation. Therefore, IBM is very useful for the computation of flow
around complex shaped, deforming and moving objects, such as swimmers. The
essence is that boundaries are represented through the introduction
of an extra force term in the NS equations by a direct-forcing scheme, while
the grid is simply Cartesian and equidistant. Velocities are
predicted on the Eulerian grid and transferred to the Lagrangian grid to
be used in the direct forcing equation. This forcing is transferred
back to the Eulerian grid to correct the predicted velocities
\cite{Fadlun2000,Iaccarino2003}. A moving-least-square approximation
builds the transfer functions from the Eulerian and Lagrangian grid
\cite{Vanella2009}. 

\noindent The number of grid cells used in these simulations is $200 \times 200
\times 240$.   The flow in the simulation is controlled through
the Reynolds number $\reyn$ and is set to $10^5$ for the hand-flow
configuration (based on hand palm width). 
%
%
At this Reynolds number the flow is turbulent. An LES (Large Eddy
Simulation) solver using a Smagorinsky model is used to model the
turbulent subgrid-scale stresses throughout the computational volume
and to mimic the small-scale motion \cite{Sagaut2002}. 
%
%
At the boundaries this LES approach results in an artificial boundary
layer, which is much thicker than the natural one. A complete
resolution of the boundary layer would have demanded ${\cal
O}(10^{11})$ grid cells on an equidistant mesh for our simulation
setup. The alternative is a locally refined mesh, but that would have
required a much more complex code. However, at current hand speeds
the boundary layer at the fingers is thin, $\delta \approx 5\times
10^{-4}$~m. This suggests that the boundary layer is not increasing
the effective frontal area of the hand significantly and is thus of minor
importance in describing enhanced drag in the finger spacing problem. 
%
%

%
\noindent The numerical method was tested on the flow around circular cylinders
and around a sphere, which are very well documented \cite{Munson2009}.
The value of the drag coefficient $C_D$ as a function of the Reynolds number was reproduced to within
10\% for $10^3 \lesssim \reyn \lesssim 10^5$, and so was the
dimensionless vortex shedding frequency for the cylinder, denoted by the Strouhal number ${\rm Sr} =
f D / U = 0.23 \pm 0.01$, with $D$ the diameter of the cylinder and
$f$ the measured frequency. As expected, the drag crisis around
$\reyn \approx 3\times 10^5$, where the upstream boundary layer turns
turbulent, is missed in these simulations due to a lack of resolution
and the used turbulence model near the surface of the body. However,
it should be noted that this value of $\reyn$ is not reached in the
simulations of the hand, in particular not around the fingers.  
%

\noindent The computational domain, shown schematically in Fig.\
\ref{fig.setup}(c), is a cuboid with size $5 \times 5 \times 6$. The
origin is located on the inlet at the center of the XY-plane. Uniform
inflow with velocity $U = 1$ was specified on the plane $z = 0.0$,
whereas radiative outflow conditions $\partial \vec{u} / \partial z =
\vec{0}$ were used at $z = 6.0$. Stress-free boundary conditions were
imposed at $y = \pm 2.5$ and no-slip conditions on the body, and periodic boundaries
were taken at $x = \pm 2.5$. 
Point $P$ (see Fig.\ \ref{fig.setup}{c}) of the $L = 3.096$ long arm
was located at $(x, y, z) = (0, 2.5, 3.0)$, which allowed a wake of
3 length units to develop. 

\noindent The wind tunnel experiments were carried out in the 8~m long and $1.1 \times
0.7 \: {\rm m}^2$ test section of a recirculation wind tunnel at the
TU/e Fluid Dynamics Laboratory. The arm was mounted on two
connected force sensors so that all three components of the total
force and one component $M_x = l_2 F_{y_2} - l_1 F_{y_1}$ of the torque could be measured.
A schematic view of the setup is shown in Fig.\ \ref{fig.setup}(b).
To give an idea of the typical magnitude of forces and their
fluctuations: at a largest mean wind velocity $U = 15.4 \; {\rm
m/s}$, and a finger spreading of $20^\circ$, the typical mean force
measured is $F_z = 6.6\; {\rm N}$, with root-mean-square (rms) fluctuation $\delta F_z =
0.3 \; {\rm N}$. For this case the Reynolds number is $\reyn =
1.03 \times 10^5$.  Due to the finite stiffness of the setup
mechanical vibrations thwarted the measurement of fluctuations due to
vortex shedding and turbulence.
Still, this measurement of the drag was preferred over one involving
the integration of measured pressures over the hand surface as its
complex geometry would have called for a prohibitive number of
pressure taps.
The forearm of the model, together with the suspension and the force
sensors partly sit in the wind tunnel boundary layer (with a typical
thickness of approximately $0.15$~m), so that the approaching flow
over the model is not entirely uniform.  Therefore, the measured
torque, which is biased towards forces on the hand, provides a better
comparison between experiment and simulations.  
%
%

\noindent The quantities of interest in both simulations and experiments
are the total forces $F_x, F_z$ on the arm and the torque $M_x$ with
respect to the point $P$ at the end of the forearm. The torque $M_x$
is interesting as it is biased towards the forces exerted on the hand
and is determined to a lesser extent by the proximal part of the arm.
Accordingly, we define drag, lift and drag-moment coefficients as
\benn 
   	C_{D, L} = \frac{F_{z, x}}{\half \rho A U^2}, 
    \;\;\mbox{and}\;\;
   	C_{M_{x}} = \frac{M_{x}}{\half \rho A L U^2},
\eenn
respectively, where $A$ is the projected area of the hand and arm perpendicular onto the direction of uniform inflow, and $L$ its length. In simulation
units, $A = 1.574$ and $L = 3.096$, which have to be scaled
accordingly for the physical arms (which have a length of 0.516~m).

\noindent The measurement of both drag and drag moment allows for a crude
distinction between the drag force $F_F$ experienced by the forearm
and the drag force $F_H$ experienced by the hand,
\benn
   	|M_x| = \half (L - H) \wt{F}_F + (L - H/2) \wt{F}_H, 
\eenn
where $H$ is the length of the hand, $H \sim 1.2$ in simulation
units, $L$ is the length of the complete hand-arm model, $L = 3.1$, and $\wt{F}_F$ and $\wt{F}_H$ are the moment-averaged forces on the arm and hand respectively, 
%
\ba
   	\wt{F}_F &=& \frac{2}{L - H} \int_0^{L - H} y \: \delta F_z(y) \: \d y, 
   	\;\; \mbox{and}\\
   	\wt{F}_H &=& \frac{1}{L-H/2} \int_{L - H}^{L} y \: \delta F_z(y) \: \d y\nn,
\ea
with $\delta F_z(y) \: \d y$ the force in the flow direction on a
slice with infinitesimal width $\d y$. For convenience, $\wt{F}_F$ is
approximated by the total force $F_F = \int_0^{L-H} \delta F_z(y) \:
\d y$, which is valid when the force across the arm is constant. 
%
%
A similar reasoning holds for $F_H$. Then
%
\ba
\label{eq.division}
   	F_H &=& \frac{2}{L} |M_x| - (1 - H/L) F_{\rm tot}, 
   	\;\;\mbox{and}\\
   	F_F &=& -\frac{2}{L} |M_x| + (2 - H/L) F_{\rm tot}\nn,   
\ea
where $F_{\rm tot} = F_F + F_H$ is the total measured force. These
are useful quantities as they can be measured both in the numerical
simulation and in the wind tunnel experiments.

\noindent Both experiments and immersed boundary simulations produced time
series of the force components $F_{x,y,z}(t)$ and the torque
$M_{x}(t)$, which allowed the computation of fluctuations. These were
quantified by their power spectra $E_{F_{x,y,z}}(f), E_{M_{x}} (f)$.
Averaged power spectra were computed using Fourier transforms over
half overlapping time windows with length $T = 73$ (simulations units), while a filter
$\sin^2(\pi t / T)$ reduced spectral leakage. Since the averages
were done over long times series, the statistical error of mean
quantities is negligible. The only inaccuracies are caused by systematic errors in the setup and method.   


\section{Results} 
\begin{figure*}[t]
\centerline{\epsfxsize = 120 mm \epsffile{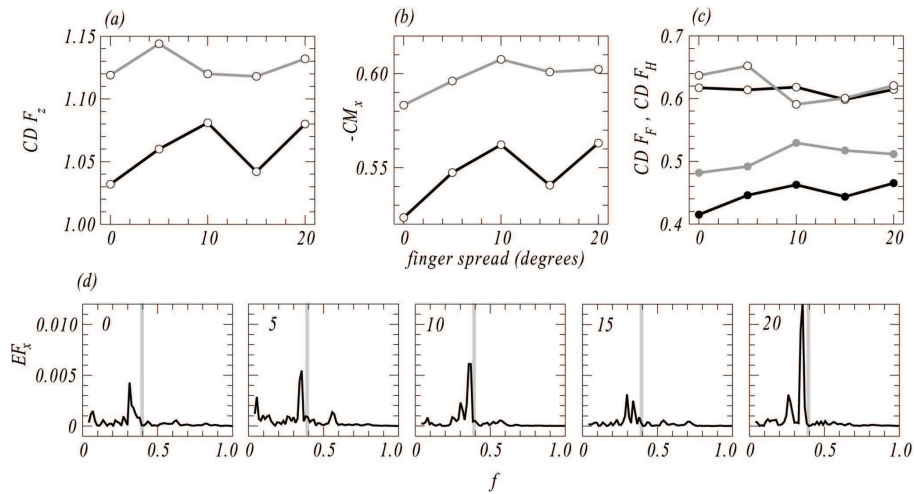}}
\caption{ 
  Panels (a, b, c) show the drag coefficient $C_D$, the drag moment 
  coefficient $C_{M_x}$ (note the minus sign) and the normalized 
  dimensionless forces on the forearm (open symbols) and the hand 
  (closed symbols). These forces are defined in Eq. \ref{eq.division}. 
  Symbols connected by black lines are numerical simulations, gray 
  lines are the experimental results. 
  Energy spectra of the dimensionless cross (lift) force $C_L$ are 
  shown in (d) for the 5 finger spreadings used. The vertical grey 
  lines indicate the computed shedding frequency of the hand and 
  fingers, assuming a Strouhal number ${\rm Sr} = 0.23$.}
\label{fig.results}
\end{figure*}

Results of the experiments and the numerical simulations are shown in Fig.\ \ref{fig.results}. The simulations predict an increase of total force and torque for every spacing larger than $0^\circ$, with local maxima around $10^\circ$ and $20^\circ$. The experiments show a similar increase for the torque, but just local maxima for the force at $5^\circ$ and $20^\circ$. 
The enhancement for the first local maximum compared to the hand with closed fingers is small, being $2\%$ and $5\%$ for the experiment and numerical simulation, respectively.  The absolute values of drag and torque differ by 10\% between experiment and simulation.  Most relevant for this study, however, is the {\em variation} of the drag and torque with finger spreading.
 

\noindent As forces on the hand have a larger weight in the evaluation of the dimensionless force moment $C_{M_x}$, the effect of optimal finger spreading is larger here, 5\% and 8\% for the experiment and numerical simulation, respectively. Also, the found optimal finger spreading of $10^\circ$ is the same in experiment and numerical simulation. The difference between the numerical and experimental results is probably due to the non-uniform flow profile on the forearm in the wind tunnel and possibly unsolved flow properties in the simulations. The second local maximum might suggest different wake structures, like the flow around a cylinder pair \cite{Sumner2010}, but should be further investigated.

\noindent Fig.\ \ref{fig.results} also shows the (approximate) division of the forces over forearm ($F_F$) and hand ($F_H$) (see Eq.\
\ref{eq.division}) in the numerical simulations and experiments.  As expected, the force on the forearm remains approximately constant, while that on the hand increases with increasing finger spreading.
Depending on the execution of the front-crawl stroke an additional
effect should be taken into account. The relative velocity of the
hand is (much) larger than that of the forearm, so that the influence
of finger spreading also becomes larger.

\noindent The energy spectra of the dimensionless side (lift) force $C_{L}$
show a peak at a dimensionless frequencies $f = 0.31 - 0.36$.  This coherent
force fluctuation is due to vortex shedding. Taking the diameter of
the forearm $0.60$ and a Strouhal number ${\rm Sr} = 0.23$, which is
characteristic for vortex shedding off a cylinder \cite{Munson2009}, the calculations predict a
frequency $f = 0.38$, which agrees well with the observed
frequency. Vortex shedding reaches a local maximum at the optimal
finger spreading, decreases at $\theta = 15^\circ$, and is strongest
at $\theta = 20^\circ$.
Both experiments and simulation show a drag minimum at $15^\circ$,
after which the drag rises again for the fingers spread uncomfortably
at $20^\circ$.  The same behavior can be observed in the vortex
shedding. It is believed that this is due to interaction between the flow around the arm and the fingers.


\section{Actuator disk model}
\begin{figure*}[t*]
\centerline{\epsfxsize = 120 mm \epsffile{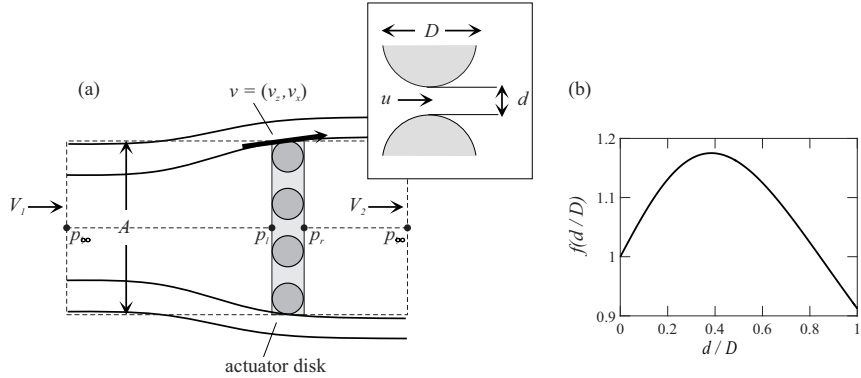}}
\caption{
   Panel (a) illustrates how to represent the hand as an ``actuator
   disk'', an effective surface of area $A$ that causes the drag.
   This actuator disk is shown greyed. The inset illustrates the flow
   $u$ in between the fingers, driven by a pressure difference
   $\Delta p$ across the hand, $u^2= \xi \Delta p d / (\rho D)$, with
   friction factor $\xi$.  The resulting normalized drag $f(d/D)$ is
   shown in panel (b).  }
\label{fig.betz}
\end{figure*}

The drag optimum of a hand with spread fingers can be described using
the actuator disk concept, which elegantly predicts the maximum
efficiency of a wind turbine \cite{Betz1966}. The idea there is to
replace the wind turbine by a disk and compute the power that passes
through the disk using Bernoulli and conservation of mass and momentum. Inspired by this concept, the four fingers with
width $D$ and gaps with width $d$ are modelled as an actuator disk
(see Fig.\ \ref{fig.betz}). A scaling description of the effect of
finger spreading is obtained by modelling the flow in between the
fingers as a pressure driven channel flow.

\noindent Briefly, let the control volume of mass- and momentum conservation be determined by the size of the actuator disk with area $A = L(4D + 3d)$. Let $V_1$ be the velocity entering the control volume and $V_2$ the mean velocity leaving the volume behind the actuator disk. Now assume, that stream tube deflection happens before the actuator disk and has a constant width behind the disk. Therefore, the mean velocity through the actuator disk $V_d$ can be approximated by $V_2$ and $p_r \approx p_{\infty}$ (see Fig.\ \ref{fig.betz}). Conservation of mass 
leaves us with an average vertical velocity $\tilde{v}_x = \frac{1}{2} (V_1 - V_2)$ at the top and bottom surfaces of the control volume, which will be used in conservation of momentum. The change in horizontal velocity from $V_1$ towards $V_2$ happens before the disk, the horizontal velocity $v_z$ can be locally approximated with the average $\frac{1}{2}(V_1 + V_2)$. Using this, conservation of momentum gives an expression for the drag force $F_D = \frac{1}{2} \rho A (V_1^2 - V_2^2)$.  
The pressure difference across the disk, $\Delta p = \frac{1}{2} \rho (V_1^2 - V_2^2)$, can be obtained with Bernoulli. Now assume, similarly to fluid flow through a channel with width $d$ and length $D$ \cite{Pope2000}, that the velocity $u$ of the water forced between the fingers follows from the pressure difference $\Delta p$ over the disk as $u^2 = \xi \Delta p \: d / (\rho D)$, with a friction factor $\xi$. Imposing mass
balance, $A \: V_2 = 3 L \: d \: u$, results in an expression for the
unknown velocity $V_2$. 

\noindent The difference with the original Betz argument is the change of
geometry of the stream tube as fingers are spread. As a function of the
dimensionless finger spreading $\alpha = d / D$ the change of the
drag, relative to that of the closed hand, is
\be
   f_D = \frac{1}{2} \rho V_1^2 4 D L \left(1 + 3 \alpha/4 \right) 
   \left( 1+ \beta \right)^{-1}
   = \frac{1}{2} \rho V_1^2 4 D L \: f(\alpha),
   \label{eq.betz}
\ee
with $\beta = (9/2) \xi \alpha^3 / (4+3\alpha)^2$.
%
The function $f(\alpha)$ is such that $f(0) = 1$. It is plotted in Fig.\
\ref{fig.betz}(b) for $\xi = 10$, which is realistic value. It shows an optimum drag enhancement
of $\sim$17.5\% for a small finger spreading $\alpha = 0.38$.  It is not very
surprising that this simple argument qualitatively predicts the drag
optimum for a small finger spreading, as it embodies the notion that
the effective area of the hand is increased by spreading the fingers
because a small opening of the fingers still presents an obstruction
of the flow between them. 
Thus, the effective area of the spread hand increases until the gap
between the fingers becomes so large that it no longer resists the
flow. Since the model is only for the four fingers, the enhancement
of the drag for the arm-hand combination is approximately one quarter
of this value. It is remarkable that this approximately agrees with
enhancement in force measured in the experiment and simulation.
%
%

\section{Conclusion}
We conclude that immersed boundary methods, which were first applied
to swimming by \citet{Loebbecke2012}, provide a
viable tool for understanding also the fine details of swimming, such
as finger spreading. The strength of IBM is its accommodation of
complex deforming boundaries at low computational costs.  
We also emphasize the usefulness of force moments which most clearly
show the influence of finger spreading. We finally stress the
importance of flow-induced force (moment) fluctuations. We find a
positive correlation between the magnitude of periodic vortex
shedding and drag, while others may have found an adverse effect
\cite{Minetti2009}.
Coherent fluid motion in the form of vortices is an example of flow
structures influencing swimming efficiency. Future research will
focus on these structures.
A scaling argument, similar to the Betz actuator disk model \cite{Betz1966}, predicts the observed drag optimum. 
%
%
Swimming propulsion necessarily comes with drag. An increase of the
drag of the hand by spreading fingers reduces the slip velocity
between the hand and the water and diminishes the power dissipated
for propulsion. This increases swimming efficiency. The effect is,
however, quite small, but may provide a competitive edge. 
%

\noindent The significance of optimal finger spreading can be roughly
quantified in terms of swimming time. A swimmer can reach a speed of
$v_s = 2$~m/s in front crawl swimming and experiences a drag force of $F_{D} = 100$~N \cite{Takagi1999,Zamparo2009,Venell2006,Novais2012}.
Assume 
that the stroke frequency of a complete cycle is 0.5~s$^{-1}$
(propulsive phase of one arm takes 1~s) and the backward
(slip) velocity of the hand with respect to the water is $v_h = 2.2$~m/s
\cite{Bilinauskaite2013,Sato1999,Kudo2013,Gourgoulis2014}.
Furthermore, assume that the projected area of the hand is
0.042~m$^2$ and that the drag coefficient $C_D = 1.03$. This yields a
propulsive force produced by the hand and arm of $F_{h} = 105$~N.  The
swimmer needs more than 200~W power to overcome drag ($P_D = F_{D} \: v_s$).
The power produced by just the hands and forearms is $P_h = F_{h} \: v_h = 230$~W (human swimming is inefficient). 
The (mechanical) power of the swimmer is given by $P_s = P_D + P_h$, with a power loss to overcome drag $P_D$ due to the motion of the complete body and a power loss due to the movement of the limbs $P_h$ for generating propulsion. Practically, the swimmer will yield the same power $P_s$, supplied muscle force ($F_h$ for the hand) and drag force $F_D$ when changing to a new technique. When obtaining an optimal finger spreading, $C_D$ can increase upto $1.08$. This allows a decrease of $v_h$ and $P_h$, since $F_h$ is fixed. The reduction of the power loss at the hand of 5~W can be used in favour of overcoming the drag $P_D$, resulting in a 2.5\% velocity increase of the swimmer $v_s = P_D / F_D$.  This equals an improvement of one's personal best of 0.6~s
on the 50~m freestyle. This result might be slightly
exaggerated. An unsteady approach would give a more underpinned
answer, as the present approach only applies to a snapshot of the
stroke. 
However, looking back at the women's 50~m freestyle final at the Rio
2016 Summer Olympics, where the first up to sixth place were decided
within 0.12~s, a marginal change in technique indeed could be the
difference between a gold medal and no medal at all.  

\section*{Conflict of Interest} 
\noindent The authors declare that they have no conflict of interest.

\section*{Acknowledgment}
This project was supported by the STW (`Stichting Technische Wetenschappen') research program Sport (project \#12868). The authors
hereby express their gratitude to Roberto Verzicco for making
available the IBM code and for help in implementing it. 

\section*{References}

\bibliography{hand} 

\end{document}